\documentclass[english,prl,aps,reprint,longbibliography]{revtex4-1}
\usepackage[T1]{fontenc}
\usepackage[latin9]{inputenc}
\usepackage{amsmath}
\usepackage{amssymb}
\usepackage{graphicx}

\makeatletter
\usepackage{babel}

\makeatother

\usepackage{babel}
\begin{document}
\title{Helical Symmetry Breaking and Quantum Anomaly in Massive Dirac Fermions}
\author{Huan-Wen Wang}
\affiliation{Department of Physics, The University of Hong Kong, Pokfulam Road,
Hong Kong, China}
\author{Bo Fu}
\affiliation{Department of Physics, The University of Hong Kong, Pokfulam Road,
Hong Kong, China}
\author{Shun-Qing Shen}
\email{sshen@hku.hk}

\affiliation{Department of Physics, The University of Hong Kong, Pokfulam Road,
Hong Kong, China}
\date{\today}
\begin{abstract}
Helical symmetry of massive Dirac fermions is broken explicitly in
the presence of electric and magnetic fields. Here we present two
equations for the divergence of helical and axial vector currents
following the Jackiw-Johnson approach to the anomaly of the neutral
axial vector current. We discover the contribution from the helical
symmetry breaking is attributed to the occupancy of the two states
at the top of the valence band and the bottom of the conduction band.
The explicit symmetry breaking fully cancels the anomalous correction
from quantum fluctuation in the band gap. The chiral anomaly can be
derived from the helical symmetry breaking. It provides an alternative
route to understand the chiral anomaly from the point of view of the
helical symmetry breaking. The pertinent physical consequences in
condensed matter are the helical magnetic effect which means a charge
current circulating at the direction of the magnetic field, and the
mass dependent positive longitudinal magnetoconductivity as a transport
signature. The discovery not only reflects anomalous magneto-transport
properties of massive Dirac materials, but also reveals the close
relation between the helical symmetry breaking and the physics of
chiral anomaly in quantum field theory and high energy physics.
\end{abstract}
\maketitle

\paragraph*{Introduction}

The chiral anomaly of massless Dirac fermions is a purely quantum
mechanical effect, and is an extraordinarily rich subject in quantum
field theory and elementary particle physics \citep{adler1969axial,Bell69nca,fujikawa1979path,Peskin-Schroeder-95book,Weinberg-quantum}.
It is regarded as a consequence of spontaneous symmetry breaking induced
by the quantum fluctuation in the presence of electric and magnetic
field. However, the helicity represents the projection of the particle
spin at the direction of motion and is also conserved for Dirac fermions.
The helical symmetry is broken explicitly in an electric field. In
the massless case the helicity and chirality become identical for
the positive energy and differ by an opposite sign for the negative
energy \citep{Dirac-quantum-mechanics,Bjoorken-book}. This raises
a question whether or not the chiral anomaly is closely related to
the explicit symmetry breaking of helicity. In recent years, the discovery
of Weyl semimetals revived research interests on chiral anomaly for
massless Dirac fermions in condensed matter physics \citep{XuS15science,Lv-15prx,hosur2013recent,armitage2018weyl,lu2017quantum,yan2017topological,bernevig2018recent}.
A negative longitudinal magnetoresistance was regarded as a significant
signature to support the existence of chiral anomaly in gapless Weyl
semimetals and Dirac semimetals \citep{Nielsen83pl,Son13prb,burkov2014chiral,kim2013dirac,XiongJ15science,zhang2016signatures,li2016negative,liang2018experimental}.
However, as more and more topological materials with finite band gap
\citep{Li16natphys,mutch2019evidence,assaf2017negative, wang2012anomalous,wiedmann2016anisotropic}
also exhibit negative magnetoresistance, the mechanism of chiral anomaly
is obviously challenged as the chiral symmetry has already been broken
explicitly by a finite mass. Very recently, Andreev and Spivak proposed
that the helicity imbalance of massive Dirac fermions may also produce
negative magnetoresistance \citep{andreev2018longitudinal}. Because
of the close relation between the helical and chiral symmetry, it
deserves an investigation on the helical symmetry breaking of massive
Dirac fermions and its transport signature in the presence of electric
and magnetic fields. Furthermore, it may reveal the deep relevance
of the helicity symmetry breaking and the physics of chiral anomaly.

In this paper, following the Jackiw-Johnson approach to the anomaly
of the neutral axial vector current \citep{jackiw1969anomalies} we
derive the equations for the divergence of the helical current and
axial vector current for massive Dirac fermions in the presence of
electric and magnetic fields. We find the discontinuity of helicity
at the momentum $q_{z}=0$ at the zeroth Landau levels leads to the
helical symmetry breaking in the presence of the electric field. The
anomalous correction from the quantum fluctuation is exactly cancelled
by the explicit symmetry breaking in the band gap. The mass term strongly
revises the coefficient in the equation for the axial vector current,
but keep the coefficient as constant in the equation of the helical
current. The two equations become equivalent in the higher energy
and massless case. The identical form of the equations for the helicity
and chirality provides deep insight into the role of helical symmetry
breaking in the physics of chiral anomaly. Physically, the helical
symmetry breaking leads to a charge current circulating along the
direction of the magnetic field, termed the helical magnetic effect.
The effect gives rise to the mass-dependent negative magnetoresistance
in massive Dirac materials.

\paragraph*{Helicity in a magnetic field}

\begin{figure}
\begin{centering}
\includegraphics[width=8cm]{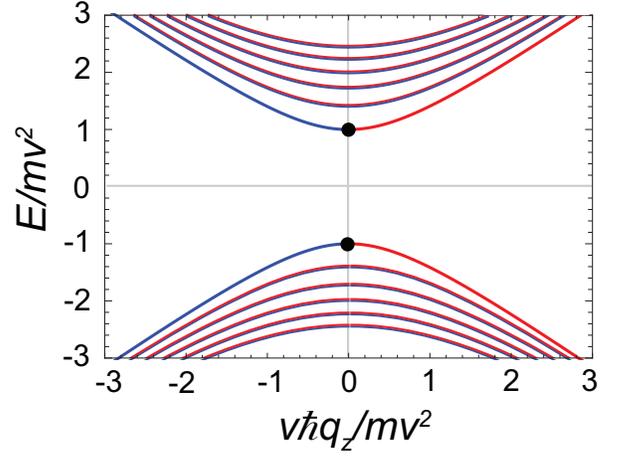}
\par\end{centering}
\caption{The energy dispersion spectrum of the Landau levels with helicity
distribution (blue line for right handed and red for left handed helicity).
The two black dots indicate the discontinuity of helicity in the Landau
levels of $n=0$. \label{fig:(a).-The-energy}}
\end{figure}

We start with the massive Dirac fermions in a finite magnetic field
$B$ along the $z-$direction,
\begin{equation}
H_{0}=\gamma^{0}\left(\gamma^{i}\Pi_{i}+mv^{2}\right)\label{eq:Dirac hamiltonian-1}
\end{equation}
where $\gamma^{0}=\tau_{1}\sigma_{0}$ and $\gamma^{i}=-i\tau_{2}\sigma_{i}$
with $\tau$ and $\sigma$ being Pauli matrices of orbital and spin
degree of freedom. $m$ is the Dirac mass and $v$ is the effective
velocity. $\Pi=\hbar\mathbf{q}+e\mathbf{A}$ is the kinematical momentum
with the vector potential $\mathbf{A}=(-By,0,0)$. Without loss of
generality, we assume $eB>0$. Since the presence of the vector potential
$\mathbf{A}$ does not break the translation symmetry along the $x$
and $z$ direction, $q_{x}$ and $q_{z}$ are still good quantum numbers.
The operator $\Sigma\cdot\Pi=\tau_{0}\sigma\cdot\Pi$ defines the
projection of the particle spin at the direction of motion. By taking
advantage of the ladder operators $a=(\Pi_{x}-i\Pi_{y})/\sqrt{2eB\hbar}$
and $a^{\dagger}=(\Pi_{x}+i\Pi_{y})/\sqrt{2eB\hbar}$ \citep{shen05prb},
we can obtain the eigenvalues and eigenstates for the operator \citep{Note-on-SM},
\begin{equation}
\sigma\cdot\Pi\left|n,q_{x},q_{z},\chi_{n}\right\rangle =\chi_{n}\sqrt{\hbar^{2}q_{z}^{2}+n\hbar^{2}\Omega^{2}/v^{2}}\left|n,q_{x},q_{z},\chi_{n}\right\rangle 
\end{equation}
where $n=0,1,2,\cdots$ are the indices of the Landau levels, and
$\Omega\equiv\sqrt{2}v\ell_{B}^{-1}$ is the cyclotron frequency with
the magnetic length $\ell_{B}=\sqrt{\hbar/eB}$ . $\chi_{n}$ stands
for the helicity of massive Dirac fermions: $\chi_{n}=\pm1$ for $n>0$
and $\chi_{0}=-\mathrm{sgn}(q_{z})$ for $n=0$. The sign change of
$\chi_{0}$ around $q_{z}=0$ is a peculiar feature of the Landau
level of $n=0$ (see the black dots in Fig. \ref{fig:(a).-The-energy}).
In the basis of helicity eigenstates, the helicity operator is expressed
as 
\begin{equation}
\hat{h}=\sum_{n,q_{x},q_{z},\chi_{n}}\chi_{n}\tau_{0}\left|n,q_{x},q_{z},\chi_{n}\right\rangle \left\langle n,q_{x},q_{z},\chi_{n}\right|.
\end{equation}
The helicity operator commutes with the Hamiltonian, $[\hat{h},H_{0}]=0$,
thus the helical symmetry survives in a finite magnetic field. In
the helicity basis, the Hamiltonian is reduced to an effective one-dimensional
system $H_{0}=\chi_{n}\sqrt{v^{2}\hbar^{2}q_{z}^{2}+n\hbar^{2}\Omega^{2}}\tau_{3}+mv^{2}\tau_{1}$.
Thus the energy eigenvalues are $\varepsilon_{n\zeta\chi_{n}}=\zeta\sqrt{v^{2}\hbar^{2}q_{z}^{2}+m^{2}v^{4}+n\hbar^{2}\Omega^{2}}$,
where $\zeta=+1$ for the conduction band and $-1$ for the valence
band. The corresponding eigenstate for each Landau level is \citep{Note-on-SM}
\begin{equation}
|n,\zeta,\chi_{n};q_{x},q_{z}\rangle=\begin{pmatrix}\cos\frac{\phi_{n\zeta\chi_{n}}}{2}\\
\zeta\sin\frac{\phi_{n\zeta\chi_{n}}}{2}
\end{pmatrix}\otimes\left|n,q_{x},q_{z},\chi_{n}\right\rangle ,\label{eq:eigenstates}
\end{equation}
where $\cos\phi_{n\zeta\chi_{n}}=\chi_{n}\sqrt{n\hbar^{2}\Omega^{2}+v^{2}\hbar^{2}q_{z}^{2}}/\varepsilon_{n\zeta\chi_{n}}$.
These eigenstates are orthogonal to each other as $\langle n',\zeta',\chi'_{n};q'_{x},q'_{z}|n,\zeta,\chi_{n};q_{x},q_{z}\rangle=\delta_{nn^{\prime}}\delta_{\zeta\zeta^{\prime}}\delta_{\chi_{n}\chi_{n}'}\delta(q_{x}^{\prime}-q_{x})\delta(q_{z}^{\prime}-q_{z})$.
All the Landau levels with different $q_{x}$ are degenerated with
the degeneracy $n_{L}=eB/2\pi\hbar$ per unit area in the x-y plane.
Besides each Landau level has additional double degeneracy for helicity
when $n>0$. 

\paragraph*{Continuity equation for helicity}

The presence of an electric field breaks the helical symmetry for
the massive Dirac fermions. Consider the electric potential $V(\mathbf{r})=e\mathbf{E}\cdot\mathbf{r}$
for a uniform electric field $\mathbf{E}$. Since the helicity operator
is a function of momentum, which does not commute the position operator
$\mathbf{r}$, $[\hat{h},\hat{V}]\ne0$. To establish the equation
of the divergence of helical currents, we follow the Jackiw-Johnson
approach to the anomaly of the neutral axial vector current \citep{jackiw1969anomalies,Peskin-Schroeder-95book},
and define the gauge-invariant helical currents 
\begin{equation}
\hat{j}_{h}^{i}(z)=\lim_{\epsilon_{\alpha}\rightarrow0}\bar{\psi}(r_{\alpha}+\frac{\epsilon_{\alpha}}{2})\gamma^{i}\hat{h}\psi(r_{\alpha}-\frac{\epsilon_{\alpha}}{2})e{}^{-i\phi(t,\epsilon_{0})}.\label{eq:current-definition}
\end{equation}
with $\phi(t,\epsilon_{0})=\int_{t-\epsilon_{0}/2}^{t+\epsilon_{0}/2}V(r_{\alpha})dt/\hbar$.
$\psi$ and $\bar{\psi}=\psi^{\dagger}\gamma^{0}$ are the Dirac spinors.
The local density and current are obtained by taking $\epsilon$ to
be small. $\hat{\rho_{h}}=\lim_{\epsilon\rightarrow0}j_{h}^{0}(z,\epsilon)$
and $\hat{j}_{h}^{i}=\lim_{\epsilon\rightarrow0}j_{h}^{i}(z,\epsilon)$.
Utilizing the time-dependent Dirac equation, the divergence of helical
currents is given by

\begin{equation}
\partial_{t}\text{\ensuremath{\rho}}_{h}+\partial_{i}j_{h}^{i}=-\frac{e^{2}}{2\pi^{2}\hbar^{2}}\mathbf{E\cdot B}-\frac{i}{\hbar}\left\langle \bar{\psi}\gamma^{0}[\hat{h},\hat{V}]\psi\right\rangle \label{eq: continuity equation_DE}
\end{equation}
where $\rho_{h}$ and $j_{h}^{i}$ are the expectation values of helical
density and current density at zero temperature. The first term in
the right-hand side of Eq. (\ref{eq: continuity equation_DE}) is
given by the anomalous correction from the quantum fluctuation 
\begin{equation}
\hat{S}(z)=i\hbar^{-1}\lim_{\epsilon_{\alpha}\rightarrow0}[\hat{j}_{h}^{0}(z,\epsilon)\epsilon_{3}-\hat{j_{h}^{3}}(z,\epsilon)\epsilon_{0}]\partial_{z}V(\mathbf{r})
\end{equation}
for small, but nonzero $\epsilon_{0}$ and $\epsilon_{3}$. The divergence
of the helical density as $1/\epsilon_{3}$ is caused by the infinity
of the Fermi sea in the valence bands, which was first encountered
in the anomaly of neutral axial vector current \citep{jackiw1969anomalies}.
Besides, the second term in the right-hand side of Eq. (\ref{eq: continuity equation_DE})
comes from the explicit helical symmetry breaking. In the basis of
the eigen energy, we find \citep{Note-on-SM} 
\begin{equation}
[\hat{h},\hat{V}]_{0}=i2eE_{z}\sum_{q_{x},q_{z}}\delta(q_{z})|0,\zeta,\chi_{0};q_{x},q_{z}\rangle\langle0,\zeta,\chi_{0};q_{x},q_{z}|
\end{equation}
for the Landau levels of $n=0$. It is noted that there exits a delta
function, which originates from the discontinuity of helicity around
$q_{z}=0$,\textit{ i.e.}, $\partial_{q_{z}}\mathrm{sgn}(q_{z})=2\delta(q_{z})$
(see Fig.\ref{fig:(a).-The-energy}). The chemical potential determines
the occupancy of the two states $\left|n=0,\zeta=\pm,\chi_{0};q_{x},q_{z}\right\rangle $
at $q_{z}=0$: each state may contribute one $\frac{e^{2}}{2\pi^{2}\hbar^{2}}\mathbf{E\cdot B}$,
which is exactly the breaking term for chiral anomaly. Thus it will
give a nonzero term $(\mathrm{sgn}(\mu)+1)\frac{e^{2}}{2\pi^{2}\hbar^{2}}\mathbf{E\cdot B}$.
Combining with the term of anomalous correction, we obtain the equation
for the divergence of the helical density $\rho_{h}$ and current
$\mathbf{j}_{h}$ in a more compact form as, 
\begin{equation}
\partial_{t}\rho_{h}+\nabla\cdot\mathbf{j}_{h}=C_{h}\frac{e^{2}}{2\pi^{2}\hbar^{2}}\mathbf{E\cdot B}.\label{eq:continuity equation-1}
\end{equation}
Here $C_{h}=\mathrm{sgn}(\mu)$ for $|\mu|\ge mv^{2}$, and $C_{h}=0$
for $|\mu|<mv^{2}$. Thus the explicit symmetry breaking term and
the anomalous correction are exactly cancelled in the right hand side
of Eq. (\ref{eq: continuity equation_DE}) within the gap. The sign
change in the conduction and valence bands is caused by the opposite
velocities of fermions with identical helicity at the direction of
the magnetic field.

\begin{figure}
\begin{centering}
\includegraphics[width=8cm]{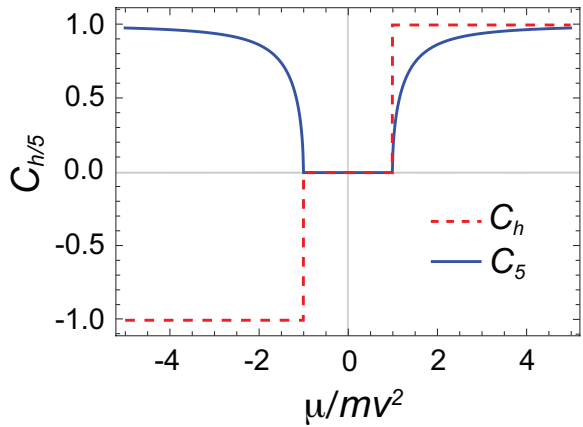}
\par\end{centering}
\caption{Comparison of the coefficients $C_{h/5}$ in the equations for the
divergence of the helical current and axial vector currents in Eqs.
(\ref{eq:continuity equation-1}) and (\ref{eq:chiral}).\label{fig:Comparison-of-the}}
\end{figure}

For the higher Landau levels of $n>0$, the diagonal elements of $[\hat{h},\hat{z}]_{n}$
always vanish and the off-diagonal elements may contribute to higher
order corrections from the electric field (see details in Ref. \citep{Note-on-SM}).
Thus Eq. (\ref{eq:continuity equation-1}) holds for a finite magnetic
field and a weak electric field. It is one of the key results of this
paper.

\paragraph*{Chiral anomaly of the massless fermions}

The helical symmetry breaking may provide an alternative approach
to derive the chiral anomaly for massless Dirac fermions. In the basis
of the eigen energy levels, the chirality operator $\gamma^{5}=i\gamma^{0}\gamma^{1}\gamma^{2}\gamma^{3}$
becomes 
\begin{equation}
\gamma^{5}=\sum_{n,\zeta,\chi_{n},q_{x},q_{z}}\zeta\chi_{n}|n,\zeta,\chi_{n};q_{x},q_{z}\rangle\langle n,\zeta,\chi_{n};q_{x},q_{z}|.
\end{equation}
For $m=0$, $\gamma_{5}$ commutes with the Hamiltonian and is conserved.
In the conduction band of $\zeta=+$1, $\gamma^{5}P_{+}=\hat{h}P_{+}$
but in the valence band of $\zeta=-$1, $\gamma^{5}P_{-}=-\hat{hP_{-}}$,
where the band projection operator $P_{\zeta}=\sum_{n,q_{x},q_{z},\chi_{n}}|n,\zeta,\chi_{n};q_{x},q_{z}\rangle\langle n,\zeta,\chi_{n};q_{x},q_{z}|$.
Thus the helicity and chirality becomes identical in the conduction
band and opposite in the valence band. Substituting the relation into
Eq. (\ref{eq:continuity equation-1}), one can obtain the continuity
equation for chiral anomaly,

\begin{equation}
\partial_{t}\rho_{5}+\nabla\cdot\mathbf{j}_{5}=\mathrm{sgn}(\left|\mu\right|)\frac{e^{2}}{2\pi^{2}\hbar^{2}}\mathbf{E\cdot B}\label{eq:chiral-anomaly}
\end{equation}
where $\rho_{5}$ and $\mathbf{\mathbf{j}_{5}}$ are the chirality
density and the corresponding axial vector current, respectively.
This provides an alternative approach to derive the chiral anomaly
from the point of view of the helical symmetry breaking. Unlike the
helicity, the chirality operator is independent of momentum, and the
presence of an electric field does not break the chiral symmetry.

\paragraph*{Pseudoscalar density and continuity equation of chirality for massive
Dirac fermions}

In the presence of a finite mass $m$, $\gamma^{5}$ does not commute
with the finite mass term, $\psi^{\dagger}[\gamma^{5},H_{0}]\psi=-2imv^{2}\mathfrak{n}_{P}$
with the pseudoscalar density $\mathfrak{n}_{P}=-\bar{\psi}i\gamma^{5}\psi$
\citep{zee2010quantum,Fang-Pseudo}. After including the contribution
from the quantum fluctuation, the divergence of the axial vector currents
is given by \citep{jackiw1969anomalies,zee2010quantum}

\begin{equation}
\partial_{t}\rho_{5}+\nabla\cdot\mathbf{j}_{5}=\frac{e^{2}}{2\pi^{2}\hbar^{2}}\mathbf{E\cdot B}-\frac{2}{\hbar}mv^{2}\left\langle n_{P}\right\rangle .\label{eq:chiral-1}
\end{equation}
In the absence of an electric field, all diagonal elements of $n_{p}$
vanish in the helicity basis, and the off-diagonal elements connect
the conduction and valence bands with the same Landau index. The expectation
value of pseudoscalar density $\left\langle \mathfrak{n}_{P}\right\rangle $
is equal to zero for a free gas of Dirac fermions. In the presence
of electric field, the electric potential may couple the two states
of the same momentum and Landau index. Due to the double degeneracy
of the states of the Landau levels of $n>0$, it is found that only
the nondegenerated Landau levels of $n=0$ contribute to the nonzero
value of $\left\langle \mathfrak{n}_{P}\right\rangle $. The perturbation
approach up to the linear electric field $\mathbf{E}$ gives 
\begin{equation}
\left\langle \mathfrak{n}_{P}\right\rangle =\left(1-C_{5}\right)\frac{e^{2}}{4\pi^{2}\hbar mv^{2}}\mathbf{E\cdot B}.
\end{equation}
The coefficient $C_{5}=\sqrt{1-\frac{m^{2}v^{4}}{\mu^{2}}}$ for $|\mu|\ge mv^{2}$
and $C_{5}=0$ otherwise \cite{Note-1}. Then, the equation for the
axial vector currents becomes \citep{Note-on-SM}

\begin{equation}
\partial_{t}\rho_{5}+\nabla\cdot\mathbf{j}_{5}=C_{5}\frac{e^{2}}{2\pi^{2}\hbar^{2}}\mathbf{E\cdot B}.\label{eq:chiral}
\end{equation}
Again the symmetry breaking term caused by the mass cancels the anomalous
correction $\frac{e^{2}}{2\pi^{2}\hbar^{2}}\mathbf{E\cdot B}$ from
quantum fluctuation out when the valence band is fully filled. Clearly,
the chirality is not a good quantum number for a nonzero mass. We
cannot define a chirality-dependent potential via the energy levels
of free fermions as we do for the helicity, but the chirality density
are still closely related to the helicity density. Consider a tiny
helical potential $\mu_{h}$ near the chemical potential $\mu$. It
is found that $\rho_{5}=\sqrt{1-\frac{m^{2}v^{4}}{\mu^{2}}}\rho_{h}$.
This demonstrates that the two quantities tend to be equal when $\mu$
is much larger than the band gap $2mv^{2}$ or the mass approaches
zero. When $\mu$ is located near the band bottom, the chirality density
approaches zero. Thus the equations for helical current and axial
current density are consistent with each other. A straightforward
comparison of the coefficients on the right hand side in Eq. (\ref{eq:continuity equation-1})
and (\ref{eq:chiral}) is presented in Fig. \ref{fig:Comparison-of-the}.
It illustrates clearly the difference and connection between the two
equations for a different chemical potential

\paragraph*{Helical magnetic effect}

For a nearly free gas of massive Dirac fermions, the helicity density
is equal to zero. If the helicity balance is broken, the chemical
potentials for fermions of different helicity deviate from the equilibrium
value $\mu$: one increases and another decreases, \textit{i.e.},
$\mu_{\pm}=\mu\pm\mu_{h}/2$. If there is no other interaction, the
helicity is still conserved. The helicity-dependent current for $\chi=\pm1$
can be calculated independently. The electric current density is given
by the difference of the helical currents for two distinct helicities
\citep{Note-on-SM},

\begin{equation}
\mathbf{j}=\frac{e^{2}}{4\pi^{2}\hbar^{2}}(\left|\mu_{+}\right|-\left|\mu_{-}\right|)\mathbf{B}.\label{eq:helical magnetic effect}
\end{equation}
This means that if two chemical potentials are not equal, a charge
current circulates at the direction of the magnetic field. We term
the field dependent current as the \emph{helical magnetic effect}
for massive Dirac fermions. The effect is equivalent to the chiral
magnetic effect when the mass $m$ approaches to zero as the the helicity
and chirality become identical \citep{vilenkin1980equilibrium,fukushima2008chiral,Huang17prb,andreev2018longitudinal}.

One remarkable transport consequence of the helical magnetic effect
is the magnetoconductivity in Dirac/Weyl semimetals. In the presence
of impurity scattering, the inter-helicity scattering process can
maintain a nonzero helical charge imbalance near the Fermi energy
$\mu$ in the background of the electromagnetic field. The scattering
potentials are functions of position, and do not commute with the
helicity operator. Thus we assume the scattering potentials $V_{s}$
are not so strong such that the averaged value of $[\hat{h},V_{s}]$
is still negligible. With a characteristic relaxation time $\tau_{\mathrm{h}}$
between different helical electrons, one can introduce a relaxation
term in the continuity equation, $\partial_{t}\rho_{h}=\mathrm{sgn}(\mu)\frac{e^{2}}{2\pi^{2}\hbar^{2}}\mathbf{E\cdot B}-\frac{\rho_{\mathrm{h}}}{\tau_{\mathrm{h}}}$.
For the equilibrium state $\partial_{t}\rho_{\mathrm{h}}=0$, the
solution for $\rho_{\mathrm{h}}$ is found as $\rho_{\mathrm{h}}=\mathrm{sgn}(\mu)\frac{e^{2}}{2\pi^{2}\hbar^{2}}\mathbf{E\cdot B}\tau_{\mathrm{h}}$
\cite{SQS}. When $|\mu_{\mathrm{h}}|\ll|\mu|$, the corresponding
helical chemical potential can be found as $\mu_{\mathrm{h}}\approx\frac{2\rho_{\mathrm{h}}}{g(\mu)}$,
where $g(\mu)$ is the total density of states at the Fermi energy.
Then, the helical magnetic effect leads to a nonzero field-dependent
current density as 
\begin{align}
\mathbf{j}_{\mathrm{HME}}= & (\frac{e}{\pi\hbar})^{4}\frac{\tau_{\mathrm{h}}}{4g(\mu)}\mathbf{E\cdot BB}.\label{eq:helical current}
\end{align}
Accordingly, the helicity-induced magnetoconductivity is given by
$\sigma_{ij}^{h}=\frac{e^{4}}{4\pi^{4}\hbar^{4}g(\mu)}\tau_{\mathrm{h}}B_{i}B_{j}.$
The inter-helicity scattering time $\tau_{\mathrm{h}}$ is determined
by the impurity scattering potentials. This equation is valid from
the weak magnetic field to the quantum limit regime. In the weak magnetic
field, the density of state at the Fermi energy is $g(\mu)=\frac{\mu q_{F}}{\pi^{2}v^{2}\hbar^{2}}$
with $q_{F}=\sqrt{\mu^{2}-m^{2}v^{4}}/v\hbar$. The matrix element
of magnetoconductivity tensor due to the helical magnetic effect reads
$\sigma_{ij}^{h}=\frac{e^{2}}{4\pi^{2}\hbar}\frac{e^{2}v^{3}}{\mu v\hbar q_{F}}\tau_{\mathrm{h}}B_{i}B_{j}$.
The longitudinal magnetoconductivity is consistent with the result
for massive and massless cases \citep{andreev2018longitudinal,burkov2014chiral,lu2017quantum}
while the transverse magnetoresistance gives rise to the planar Hall
effect \citep{LiH-18prb}. In the Born approximation, the inter-helicity
scattering time is found to reach a maximal value at $m=0$, and decays
with the increase of the mass, which results in a mass-dependent magnetoconductance
in Dirac materials \cite{Note-on-SM}. In the quantum limit regime,
the density of state at fermi energy $\mu$ is found as $g(\mu)=\frac{\mu}{2\pi^{2}\ell_{B}^{2}v^{2}\hbar^{2}q_{F}},$
and the corresponding longitudinal magnetoconductivity becomes $\sigma_{zz}^{\mathrm{h}}=\frac{e^{3}v}{2\pi^{2}\hbar^{2}}\frac{v\hbar q_{F}}{\mu}\tau_{\mathrm{h}}B$,
which is a linear function of the magnetic field once $\tau_{h}$
is a constant in the massless limit. For a moderate strong magnetic
field, the density of states will oscillate with the magnetic field,
and there are quantum oscillations in the magnetoconductivity.

\paragraph*{Discussion and conclusion}

The full cancellation of the explicit symmetry breaking and the anomalous
correction in the band gap reflects the quantum anomaly in the massive
Dirac fermions. The anomalous correction arises by introducing the
gauge invariant currents in Eq. \ref{eq:current-definition}, and
is an electromagnetic response from the infinite Dirac sea. Here it
is worth of pointing out that the approach is different from the method
based on the variation of charge density of particles in the zeroth
Landau levels \cite{Nielsen83pl,andreev2018longitudinal}, in which
all other negative energy levels are neglected and is actually irrelevant
to physics of the quantum anomaly. This point can be further clarified
in the following example. Consider the nonrelativistic Pauli Hamiltonian
for a free electron gas in a magnetic field, $H_{P}=\left(\sigma\cdot\Pi\right)^{2}/2m=\Pi^{2}/2m+\frac{e\hbar}{2m}B\sigma_{z}$.
The helicity is conserved, $[\sigma\cdot\Pi,H_{P}]=0$. The helical
symmetry breaking in an electric field leads to an identical continuity
equation for the divergence of helical density and current as in Eq.
(\ref{eq:continuity equation-1}) for $\mu>0$ \cite{Note-on-SM},
which is consistent with the picture of the Landau levels. However,
it is unrelated to the physics of quantum anomaly since there are
no infinites negative energy states at all. The helical symmetry breaking
in this system may also give rise to a negative longitudinal magnetoresistance.
Of course it should be noted that the effect disappears if the Zeeman
field is absent, i.e., in $H=\Pi^{2}/2m$.

In summary, we derived the two equations for the divergence of helical
current and axial vector current in electric and magnetic fields.
We discovered the discontinuity of helicity at the zeroth Landau levels
leads to the helical symmetry breaking in the presence of the electric
field. The occupancy of the states at the top of the valence band
and the bottom of the conduction band contributes one in the unit
of $\frac{e^{2}}{2\pi^{2}\hbar^{2}}\mathbf{E\cdot B}$ in the equation
of the divergence of the helical currents. The anomalous corrections
from the quantum fluctuation for both helicity and chirality are cancelled
exactly by the explicit symmetry breaking to guarantee the conservation
laws when the chemical potential is located within the energy gap.
In the case of higher energy or tiny mass $\left|\mu\right|\gg mv^{2}$,
the equations for helicity and chirality become equivalent (only differed
by a sign for positive and negative energy). This provides an alternative
route to understand the chiral anomaly from the point of view of the
helical symmetry breaking. The two equations may shed some new insights
to the physics of the chiral anomaly in the quantum field theory as
well as peculiar transport behaviors in condensed matter. For instance,
as a peculiar feature of massive Dirac fermions in a magnetic field,
the helical magnetic effect can gives rise to a mass-dependent positive
magnetoconductance.
\begin{acknowledgments}
This work was supported by the Research Grants Council, University
Grants Committee, Hong Kong under Grant No. 17301220, and the National
Key R\&D Program of China under Grant No. 2019YFA0308603. 
\end{acknowledgments}

\bibliographystyle{apsrev}


\end{document}